\documentclass[aps,prl,twocolumn,floatfix,amssymb]{revtex4-1} 
\usepackage[utf8]{inputenc}
\usepackage[english]{babel}
\usepackage{amsmath}
\usepackage{amsfonts}
\usepackage{amssymb}
\usepackage{graphicx}
\usepackage{float}
\usepackage{color}
\usepackage{braket}
\usepackage{sidecap}
\usepackage{mciteplus}
\usepackage[breaklinks=true,colorlinks,citecolor=blue,linkcolor=blue,urlcolor=blue]{hyperref}
\newcommand{\be}{\begin{equation}}
\newcommand{\ee}{\end{equation}}
\newcommand{\ber}{\begin{eqnarray}}
\newcommand{\eer}{\end{eqnarray}}

\newcommand{\rv}{{\bf r}}

\def\eer{\end{eqnarray}}

\def\rv{{\bf r}}

\begin{document}

\title{Origin of the Temperature Collapse of the Electric Conductivity in Bilayer Graphene}
\author{Mohammad Zarenia$^1$, Shaffique Adam$^{2,3,4}$, and Giovanni Vignale$^{1,2,3}$}
\affiliation{$^1$Department of Physics and Astronomy, University of Missouri, Columbia, Missouri 65211, USA\\
$^2$Yale-NUS College, 16 College Ave West, 138527 Singapore\\
$^3$Centre for Advanced 2D Materials, National
University of Singapore, 6 Science Drive 2, 117546, Singapore\\
$^4$Department of Physics, National
University of Singapore, 2 Science Drive 3, 117551, Singapore}

\begin{abstract}
Recent experiments have reported evidence of dominant electron-hole scattering in the electric conductivity of suspended bilayer graphene near charge neutrality. According to these experiments, plots of the electric conductivity as a function of $\mu/k_BT$ (chemical potential scaled with temperature) obtained for different temperatures in the range of $12\rm{K}\lesssim T \lesssim 40\rm{K}$ collapse on a single curve independent of $T$.  In a recent theory, this observation has been taken as an indication that the main sub-dominant scattering process is not electron-impurity but electron-phonon.  Here we demonstrate that the collapse of the data on a single curve can be explained without invoking electron-phonon scattering,  but assuming 
that the suspended bilayer graphene is not a truly gapless system. 
With a gap of $\sim 5$ meV, our theory produces excellent agreement with the observed conductivity over the full reported range of temperatures.  These results are based on the hydrodynamic theory of conductivity, which thus emerges as a solid foundation for the analysis of experiments and the estimation of the band-gap in multiband systems.
\end{abstract}
\maketitle
{\it Introduction.} In recent years, the investigation of  intrinsic transport properties of two-dimensional (2D) systems has been the subject of intense interest
\cite{bandurin, ghahari, crossno, Principi_prb_2016,Narozhny_prb_2015,Briskot_prb_2015,Fritz_2008,Muller_2008, foster,svintsov2013,svintsov2018,shaffique, Principi_conductivity,lucasWF,lucasKT}. 
Bilayer graphene (BLG) is one of the most carefully scrutinized systems within this class.  With a quadratic low energy dispersion around the Fermi level, extremely high purity,  and the possibility of opening a tunable gap between the conduction and valence bands by applying a potential difference between the gates, it is considered a promising platform for electronic devices.~\cite{McCann2013, Zhang2009}.

Recent experiments have demonstrated the dominant role of momentum-conserving electron-hole (e-h) scattering in defining the electric conductivity of ultra-clean BLG systems \cite{ Morpurgo2017,tan2019}.  The dominance of momentum conserving interactions defines a ``hydrodynamic regime".  Transport in this regime is quite different from conventional single-particle transport in which the conductivity is defined by momentum non-conserving interactions, such as electron-impurity and electron-phonon collisions.  Ref. \cite{Morpurgo2017} presents measurements of the conductivity around the charge neutrality in nominally zero-gapped suspended BLG samples. More recent work \cite{tan2019} reports the observation of  hydrodynamic transport in bias-induced gapped BLG  encapsulated in hexagonal boron nitride substrates, at the charge neutrality point. 
In the hydrodynamic regime, where e-h collisions are dominant, the conductivity of single and double layer graphene near the charge neutrality point (i.e., for $\mu\ll k_BT$, where $\mu$ is the chemical potential, which vanishes at the charge neutrality point), is described by the following expression~\cite{zarenia,zareniaBLG,zareniaTwisted}, 
\be\label{eqS}
\frac{\sigma(\mu,T)}{\sigma_{0}(T)}\simeq 1+\frac{2~\bar\mu^2}{\Gamma(T)^2},~~\bar\mu=\mu/(k_BT)\, 
\ee
where  $\sigma_0(T)$ is the intrinsic conductivity due to the e-h scattering at the charge-neutrality $\mu=0$ (first calculated for monolayer graphene in Refs.~\cite{kashuba,sachdev}).   The crucial quantity
\be\label{Gamma}
\Gamma(T)=\sqrt{\frac{\sigma_{0}(T)}{\sigma_{\rm{dis}}(T)}}
\ee
is the ratio between the intrinsic conductivity $\sigma_0(T)$, calculated by taking into account {\it only} the momentum-conserving electron-hole interactions, and the disorder conductivity  $\sigma_{\rm{dis}}(T)$ calculated by taking into account {\it only} the sub-dominant momentum-non conserving collision of electrons (or holes) with impurities and phonons under the same conditions ($\bar \mu=0$).   Obviously, the hydrodynamic regime is realized only if $\Gamma\ll 1$, i.e., if electron-hole collisions are much more frequent than electron (hole)-impurity or electron (hole)-phonon collisions.  Notice that neither $\sigma_0$ nor $\sigma_{\rm dis}$ is the true physical conductivity.  In fact, Eq.~(\ref{eqS}) defines a highly unusual regime of conduction in which the electron-hole and the electron-impurity scattering mechanisms compete against each other like resistors connected in parallel, rather than adding up their contributions like resistors connected in series (the so-called Matthiessen's rule).

Remarkably, Eq.~(\ref{eqS}) has been independently obtained~\cite{simon} by using a two-fluid model, in which electrons and holes fluids respond to an external electric field, scatter independently from impurities and/or phonons, and from each other via their mutual Coulomb interaction.  The beauty of the equation lies in the fact that the cumulative effect of all types of disorder, e.g. charged impurities, phonons, etc., is included in a single parameter  $\Gamma(T)$ through the collision-limited conductivity $\sigma_{\rm{dis}}(T)$ (at low doping level around the charge neutrality, the dependence of $\sigma_{\rm{dis}}(T)$ upon doping level can be  neglected).   Simply stated, $\Gamma^{-1}$ determines the curvature of the conductivity plotted as a function of scaled chemical potential $\bar \mu$.  The rapid  rise of  $\sigma$ as a function of $\bar \mu$ reflects the activation of the center of mass mode of the electron-hole system, which becomes electrically active as soon as $\bar \mu \neq 0$.\cite{zarenia,zareniaBLG} 

The experiments of Ref.~\cite{Morpurgo2017} and~\cite{tan2019} confirm the quadratic dependence of $\sigma$ on $\bar \mu$, but, in addition, provide important information about the behavior of $\Gamma$ as a function of temperature.  Specifically, Ref.~\cite{tan2019} shows that plots of the conductivity vs temperature for different values of the bias-induced gap $\Delta$ at $\mu=0$ collapse onto a single curve when plotted as functions of $\bar\Delta=\Delta/k_BT$.  This can be understood by noting that in the presence of a finite gap, $\Delta$, both the intrinsic conductivity $\sigma_{\rm{CNP}}(T)$ and the disorder conductivity  $\sigma_{\rm{dis}}(T)$  have a gap dependence that scales with $\Delta/k_BT$.   This is because the conductivities are proportional to the density of thermally excited carriers multiplied by an appropriate scattering time (electron-hole, electron-impurity, electron-phonon).  The scattering times are independent of $\Delta$, as the corresponding scattering processes do not involve transfer of carriers between the conduction and the valence band. On the other hand, the densities of thermally excited carriers at $\mu=0$, scale as $\Delta/(k_BT)$.  Therefore $\Gamma$ and the whole $\mu=0$ conductivity depends only on the ratio $\Delta/(k_BT)$, showing that Eq.~(\ref{eqS}) is consistent with the data.

The data of Ref.~\cite{Morpurgo2017} for finite, but small values of $\bar \mu$ are more puzzling. Plots of the conductivity  as a function of  $\bar\mu$  in nominally gapless BLG ($\Delta=0$)  for different temperatures are found to collapse on a single parabola $\propto \bar \mu^2$, with a curvature independent of temperature in the range  $12\rm{K}\lesssim T\lesssim 40\rm{K}$.   According to Eq.~(\ref{eqS}), these data imply that $\Gamma(T)$ is independent of $T$ in the stated range of temperatures. 
It is well known that, in the absence of a gap,  $\sigma_0$ is very weakly (logarithmically) dependent on temperature: this comes about because the density of thermally excited carriers scales as $T$, while the electron-hole scattering time scales as $1/T$ (the weak logarithmic dependence  arises from the renormalization group flow of the electron-hole coupling constant~\cite{Fritz_2008}, and will be neglected from now on).  Consider now the temperature dependence of $\sigma_{\rm{dis}}(T)$.  This depends crucially on the subdominant scattering mechanism.  For the electron-impurity collisions, the scattering time is expected to be independent of $T$,  yielding   $\sigma_{\rm{dis}}(T) \propto T$.  This would give $\Gamma(T) \propto T^{-1/2}$ in glaring contradiction with the experimental data.  On the other hand,  for  electron-phonon collisions, the scattering time is expected to scale as $1/T$, yielding a temperature-independent $\sigma_{\rm{dis}}$ and an essentially temperature-independent $\Gamma$, in agreement with the experimental data.  Drawing attention to this fact, Weber {\it et al.}~\cite{simon}  have recently argued that the main momentum-non-conserving interaction for the experiments of Ref.~\cite{Morpurgo2017} is not electron-impurity, but electron-phonon. 
%

This conclusion runs counter to the prevailing opinion that electron-phonon scattering  in BLG is relevant only for $T\gtrsim 100$K (see Refs. \cite{Guinea2011,Hakonen2015}),  whereas  the experimental observation of ``conductivity collapse"  happens at much lower temperatures ($T\lesssim 50$K). 

In this paper, we show that the experimental observations of Ref.~\cite{Morpurgo2017} can be explained in
quantitative detail within the conventional framework of electron-impurity scattering, without involving phonons, provided  we assume that the suspended   BLG samples are not truly gapless systems, but have a finite gap on the order of a few meV.  Specifically, we show that in the presence of a small temperature-independent gap, the intrinsic Coulomb conductivity $\sigma_0$ acquires a temperature dependence (scaled with $\Delta/k_BT$ as shown in Ref. \cite{tan2019}), which is nearly exactly linear and precisely cancels the temperature dependence of the impurity-limited conductivity in the range of experimental temperatures.  With a gap of $\sim 5$ meV, our theory produces excellent agreement with the observed conductivity data. Thus, our analysis of the experimental data provides concrete evidence for the existence of a finite gap.\\

\begin{figure}[t]
\includegraphics[width=9cm]{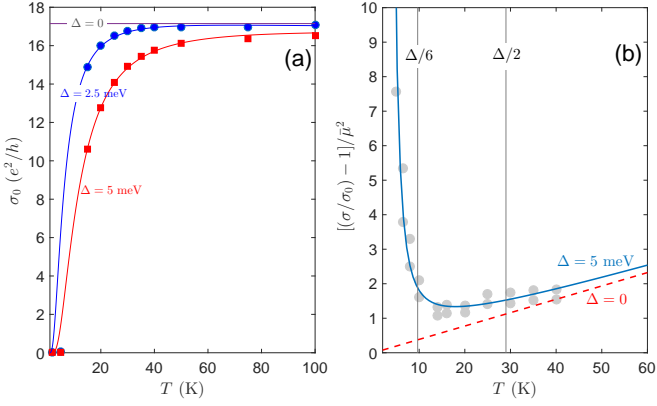}
\caption{
(a) Intrinsic electron-hole conductivity $\sigma_0$ at charge neutrality in  BLG as a function of temperature for $\Delta=0$ (gapless BLG), $\Delta=2.5$ meV, and  $\Delta=5$ meV, as labeled. Symbols are the numerical results obtained by solving the quantum Boltzmann equation and the solid curves are the analytic results obtained from Eq.~ (\ref{eqSCNP}). 
While $\sigma_0 (\Delta=0)$ is independent of temperature, $\sigma_0 (\Delta\neq0)$ acquires a temperature dependence in the presence of an energy gap. (b)
Electric conductivity $[(\sigma(\bar\mu,T)/\sigma_0)-1]/\bar\mu^2$ (i.e. $2/\Gamma^2$ in Eq. (1), as a function temperature. Gray dots are the experimental data of Device-3 in Ref. \cite{Morpurgo2017} for $\bar\mu=0.5$ and $\bar\mu=0.75$. The blue solid curve is the theoretical result obtained using Eq. (\ref{eqS}) for the finite gap of $\Delta=5$ meV. The red dashed line shows the results for $\Delta=0$. The charged-impurity density $n_{\rm{imp}}=1\times10^{10}$ cm$^{-2}$, and short-range potential $V_0=2300$ meV.nm$^2$.  In the presence of a small finite gap, we find excellent agreement with experiments over the full range of the reported temperatures. The temperature collapse occurs for $\Delta/6\lesssim T\lesssim \Delta/2$, as indicated by the vertical lines.}
\label{fig1}
\end{figure}

{\it Theory.} The intrinsic $\sigma_0$ has been calculated for gapless clean BLG by numerically solving the e-h Coulomb collisions within the quantum Boltzmann equation with the help of the Fermi golden rule (see the details of the calculations  in  \cite{zareniaBLG}). For zero-gapped BLG ($\Delta=0$), $\sigma_0$ is nearly independent of $T$ (the gray line in Fig.~\ref{fig1}a). In the presence of an energy gap $\Delta$, however, we find that $\sigma_0$ acquires a $T$-dependence. The symbols in Fig.~\ref{fig1} show the numerical results for $\sigma_0$ obtained from the solution of the quantum Boltzmann equation for $\Delta=2.5$ meV and  $\Delta=5$ meV, as labeled.  At low temperatures, $T\ll\Delta$, the gap suppresses the conductivity. $\sigma_0(T,\Delta)$ increases with temperature and approaches the conductivity of gapless BLG $\sigma_0(\Delta=0)$ at  $T\gg\Delta$, where the gap is no longer relevant. For suspended BLG, we obtain  $\sigma_0(\Delta=0)\approx 17e^2/h$. 

To find a convenient analytic expression for the $T$-dependence of  charge neutral  e-h conductivity in gapped BLG, we employ a Drude model and write  
$\sigma_{0}(\Delta,T)=n(\Delta, T) e^2\tau_{\rm{eh}}/m^\ast$,
where $n(\Delta,T)$ is the density of thermally excited carriers, $\tau_{\rm{eh}}$ is the e-h quasi-particle life time, and $m^\ast$ is the effective mass of the gapless parabolic bands. Assuming that $\tau_{\rm{eh}}$ remains unchanged in the presence of a small gap (quite a reasonable assumption given that the electron-hole scattering process does not transfer carriers between bands)  we obtain
\be
\sigma_0(\Delta\neq 0,T)\sim \sigma_0(\Delta=0)~ \frac{n(\Delta,T)}{n(0,T)},
\ee
where $\sigma_0(\Delta=0)$ is the e-h conductivity of gapless BLG, shown by the gray line in Fig. \ref{fig1}. The carrier density as a function of $\Delta$ and  $T$ is given by 
\be
n(\Delta,T)=\int_\Delta^\infty N(\epsilon)f(\epsilon) d\epsilon,~N(\epsilon)=\frac{N_0|\epsilon|~\theta(|\epsilon|-\Delta)}{\sqrt{\epsilon^2+(\Delta/2)^2}}\,,
\label{eqn}
\ee
where $ f(\epsilon)=[\exp{(\epsilon/k_BT)}+1]^{-1}$ is the Fermi-Dirac distribution at charge-neutrality ($\mu=0$) and $N(\epsilon)$
is the density of states corresponding to the hyperbolic bands of gapped BLG.  The constant $N_0 = {m^\ast}/{(\pi\hbar^2)}$ is the density of states of the parabolic bands of gapless BLG.


\begin{figure}
\hspace{-1cm}
\includegraphics[width=10cm]{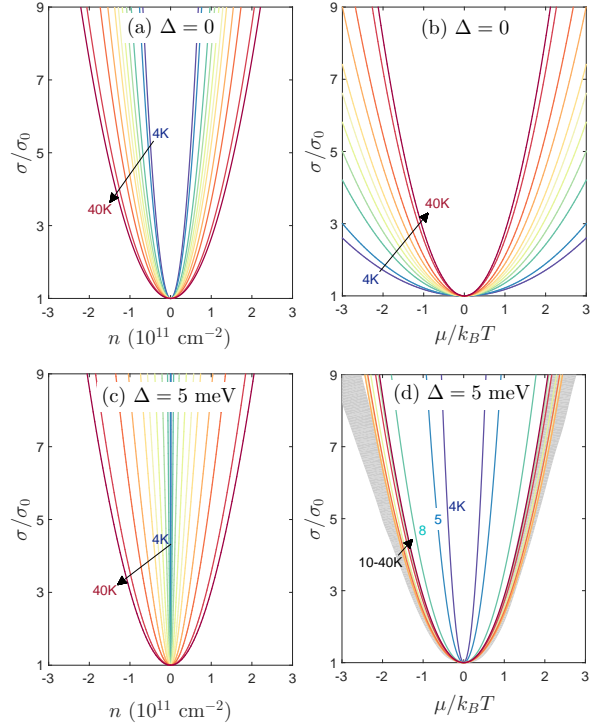}
\caption{Electric conductivity $\sigma(n,T)$ as a function of density (a,c) and the corresponding $\mu/k_BT$ (b,d) for gapless BLG (a,b) and gapped BLG with $\Delta=5$ meV (c,d) at various temperatures, $T=4,~5,~8,~10,~15,~20,~25,~30,~35,~40$ K as indicated ($T$ increases from blue to red).  The charged impurity density is $n_{\rm{imp}}=1\times10^{10}$ cm$^{-2}$ and the short-range potential is $V_0=2300$ meV.nm$^2$.
As functions of doping concentration $n$, the conductivity plots are well separated, no matter whether BLG is gapless or has a finite gap. As functions of $\mu/k_BT$, plots of the conductivity for different temperatures collapse onto a single curve only if a gap is present. In the presence of a finite gap, the conductivity-$\mu/k_BT$ plots (d),  decrease for $T=4,5,8$ K, stabilize for $10~\rm{K}\leq T\leq 30$ K, and start to increase when $T\gtrsim 40$ K. The gray shaded region in (d) shows the experimental data of Ref. \cite{Morpurgo2017} for $12~\rm{K}\leqslant T\leqslant40$ K.}\label{fig2}
\end{figure}

For small $\Delta$ ($\Delta \ll k_BT$) the carrier density can be approximated as $n(\Delta,T)\approx (1+\Delta/2k_BT)$ and thus the intrinsic e-h conductivity takes the form

\be
\sigma_0(\Delta,T)\sim \sigma_0(\Delta=0)\exp{\left(-\frac{\Delta}{2k_BT}\right)} \left(1+\frac{\Delta}{2k_BT}\right). 
\label{eqSCNP}
\ee

%

We now assume, as it is commonly done, that the primary mechanism for momentum relaxation in the temperature range covered by the experiment of Ref.~\cite{Morpurgo2017} is scattering from charged impurities, i.e.,  $\sigma_{\rm{dis}}=\sigma_{\rm{e-imp}}$.
Using a simple model of electrons and holes scattering against randomly distributed impurities of density $n_{\rm{imp}}$ with short-range potential $V_0\delta(\rv)$, the impurity-limited conductivity of the system is found to be given by  $\sigma_{\rm{e-imp}}=n e^2\tau_{\rm{e-imp}}/m^\ast$, where the $\tau_{\rm{e-imp}}=\hbar^3/(m^\ast n_d V_0^2)$ is temperature-independent for parabolic bands and fixed by the impurity strength $n_{\rm{imp}}V_0^2$. At the charge neutrality point $n=2m^\ast k_BT/\pi\hbar^2$ and thus we have $\sigma_{\rm{e-imp}}\sim T$. In our calculations we assume typical values of $n_{\rm{imp}}=1\times10^4$ cm$^{-2}$, and $V_0=2300$ meV.nm$^2$.  Including the effect of screening (long-ranged charged impurities) does not change the linear $T$-dependence of $\sigma_{\rm{e-imp}}$ (see the supplementary information in Ref. \cite{tan2019}).

In order to define a temperature range for which $\sigma_0(\Delta,T)$ behaves linearly with $T$, we calculate  the inflection point of  Eq. (\ref{eqSCNP}) as a function of $T$, and find $T_{\rm inflection}=\Delta/2k_B$.  From the slope of the function at the inflection point we obtain that for  $\Delta/6\lesssim k_BT\lesssim \Delta/2$, the $T$-dependence of $\sigma(\Delta,T)$ is approximately linear and compensates the linear $T$-dependence of $\sigma_{\rm{e-imp}}$ in Eq.~(\ref{Gamma})  for $\Gamma(T)$. 

In Fig. \ref{fig1}b, we compare our theoretical results for $\sigma(n,T)$ with the experimental data of Ref. \cite{Morpurgo2017}. We obtain excellent agreement with experiments by taking a constant gate-induced $\Delta=5$ meV. For $(\Delta/6\approx10\rm{K})\lesssim T\lesssim (\Delta/2\approx30\rm{K})$, the values $\sigma(\mu,T)$ become nearly independent of $T$, as observed in experiments.
The dashed lines in Fig.~\ref{fig1}b show the conductivity calculated for zero-gapped BLG.   The zero-gap conductivity keeps increasing linearly with temperature throughout the range of the experiment. 
While the available experimental data were limited to temperatures $T\lesssim40$K,  we expect that $\sigma(n,T)$ would begin to increase linearly as a function temperature for $T>\Delta/2$, where the gap begins to be irrelevant. 

Lastly, Fig.~\ref{fig2} shows  $\sigma(n,T)$ plotted as a function of doping concentration $n=n_e-n_h$ (left panels), where $n_e$ and $n_h$ are the densities of electrons and holes, and $\mu/k_BT$ (right panels) for different temperatures as indicated in the figure.  The upper panels are for gapless BLG ($\Delta=0$) and the lower panels are for gapped BLG with $\Delta=5$ meV.  As functions of doping concentration, the conductivity plots are well separated, no matter whether BLG is gapless or has a finite gap (see Figs.~\ref{fig2}a and \ref{fig2}c). As functions of $\mu/k_BT$, plots of the conductivity for different temperatures collapse onto a single curve only if a gap is present (compare Figs.~\ref{fig2}b and \ref{fig2}d).  Notice that in Fig.~\ref{fig2}d,  consistent with the results shown in Fig.~\ref{fig1}b,  the conductivity  plots show temperature dependence for $T=4,5,8$K, stabilize for $10\rm{K}\leq T\leq 30$ K, and start changing again when the temperature grows above $\sim 40$K. These results are in excellent agreement with experiments. The gray shaded region in Fig.~\ref{fig2}d shows the experimental data of Ref. \cite{Morpurgo2017} (device 1) for $12~\rm{K}\leqslant T\leqslant 40$ K.

{\it Dicussion} --   Our findings are in contrast with the outcome of a ``Planckian analysis" from which the authors of Ref. \cite{Morpurgo2017} extract the gap of their BLG samples in a more recent experiment~\cite{Nam2018}. In this study, the authors assume that the electric conductivity away from charge neutrality is given by a Drude formula with the carrier densities of the gapless BLG model and a mean free time equal to the electron-hole scattering time. 
However, there is a fundamental problem with this assumption.  Use of the Drude formula fails to account for the emergence, {\it away from charge neutrality},  of  a resistance-free channel of conduction  -- the center of mass momentum mode -- which would cause the conductivity to diverge in the absence of momentum-non-conserving collisions.  Therefore the dependence of the conductivity upon doping is severely distorted by the Drude modeling.
Furthermore,  we have found that the Planckian Drude formula~\cite{Planckian} when fitted to the conductivity data for $12\rm{K}\lesssim T\lesssim 40$K  produces a linearly  $T$-dependent  gap,  $\Delta \sim 2k_BT$ for the samples in Refs. \cite{Morpurgo2017}, which is incompatible with any known physical mechanism for the generation of a gap.

{\it Conclusion.} We have proposed a theoretical explanation for the recent experimental reports of the ``temperature collapse" of electric conductivity in BLG. Experiments in suspended BLG samples show that around the charge neutrality and  in the electron-hole scattering dominated transport regime, plots of the electric conductivity as a function of $\mu/k_BT$ (chemical potential scaled with temperature) collapse on a single temperature-independent curve for temperatures  $12\rm{K}\lesssim T \lesssim 40\rm{K}$ \cite{Morpurgo2017}. In contrast with a recent theoretical suggestion we propose that the main momentum-non-conserving scattering mechanism  can be electron-impurity (and hole-impurity) scattering.  This proposal is viable if we assume that suspended BLG is not a truly gapless system, as suggested by previous experiments~\cite{Freitag2012,Velasco2012,Weitz2010,Oostinga2008} and theories \cite{Nomura2006, Min2008, Zhang2010, Kharitonov2012, Nandkishore2010, Koshino2017}. With a small gap $\Delta \simeq 5$ meV, the intrinsic conductivity at charge neutrality acquires  a temperature dependence which compensates for the $T$-dependence of the impurity-limited conductivity.  We find excellent agreement with the experimental results in the temperature range  $4\rm{K}\lesssim T\lesssim 40$K, without invoking the electron-phonon interaction.  At higher temperatures, where the electron-phonon scatterings become relevant, i.e. $T\gtrsim 100$K in BLG, the intrinsic gap is no longer important ($T\gg \Delta$), and the theoretical analysis of Ref.~\cite{simon}  should explain the behavior of the conductivity.  Thus, our analysis rescues  electron- (hole-) impurity collisions as a viable mechanism for modeling the conductivity of BLG near the charge neutrality point, and provides evidence for the existence of an intrinsic gap in this nominally gapless system.  Beyond the particular experimental data we have considered here, we have shown that the hydrodynamic theory of conductivity, Eq. (1), provides a sound foundation for the estimation of the band gap in future experimental samples of multiband systems.
\begin{acknowledgements}
{\it Acknowledgment.} This work was supported by the U.S. Department of Energy (Office of Science) under grant
No. DE-FG02-05ER46203.  We would like to thank  Youngwoo Nam, Dong-Keun Ki, David Soler-Delgado, and  Alberto Morpurgo for providing us with their experimental data of Ref. \cite{Morpurgo2017} .
\end{acknowledgements}

%

\end{document}